\documentclass[aps,pra,twocolumn,10pt,superscriptaddress,nofootinbib]{revtex4-1} 

\usepackage{amsmath}

\usepackage{graphicx}
\usepackage{amsfonts}
\usepackage{braket}
\usepackage{natbib}
\usepackage{amssymb}
\usepackage[utf8]{inputenc}
\usepackage{dsfont}
\usepackage{xcolor}
\usepackage{subfigure}
\usepackage[colorlinks=true,citecolor=blue,linkcolor=blue]{hyperref}

\newcommand{\LR}[1]{\left(#1\right)}
\newcommand{\Trace}[1]{\text{Tr}\left\{#1\right\}}		

\begin{document}

\title{Optimized controlled Z gates for two superconducting qubits coupled through a resonator}

\author{D. J. Egger}
\affiliation{Theoretical Physics, Universit\"{a}t des Saarlandes, 66123 Saarbr\"{u}cken, Germany}
\author{F. K. Wilhelm}
\affiliation{Theoretical Physics, Universit\"{a}t des Saarlandes, 66123 Saarbr\"{u}cken, Germany}
\affiliation{IQC and Department of Physics and Astronomy, University of Waterloo, Ontario N2L 3G1, Canada}

\date{\today}

\begin{abstract}
Superconducting qubits are a promising candidate for building a quantum computer. A continued challenge for fast yet accurate gates to minimize the effects of decoherence.  Here we apply numerical methods to design fast entangling gates, specifically the controlled Z, in an architecture where two qubits are coupled via a resonator. We find that the gates can be sped up by a factor of two and reach any target fidelity. We also discuss how systematic errors arising from experimental conditions affect the pulses and how to remedy them, providing a strategy for the experimental implementation of our results. We discuss the shape of the pulses, their spectrum and symmetry.
\end{abstract}

\maketitle

\section{Introduction}

Superconducting quantum devices provide a promising route to creating a quantum computer \cite{Schoelkopf_Nature_451_664}. In many applications, the quantum states are implemented with qubits \cite{Clark_Nature_453_1031} connected by strip-line resonators \cite{Blais_PRA_69_062320}. The qubit is engineered to have a strong dipole interaction with the cavity. This strong coupling allows many Rabi oscillations between qubits and resonators before the quantum states decohere. Although coherence times have significantly improved over the past decade  
\cite{Nakamura99,Vion02,Bertet05,Houck09,Bylander_NatPhys_7_565,Paik_PRL_107_240501,Masluk12}
the quantum operations should be implemented quickly to mitigate the effects of decoherence. Additionally if a full scale quantum computer were to be built many quantum operations have to be performed and thus gate speed is crucial to limit computation times \cite{Stock09}. Human-engineered artificial atoms have great flexibility and controllability \cite{Insight,Schoelkopf_Nature_451_664,Makhlin01,Devoret04,Devoret13}. Therefore there is much to be gained by using optimal control theory methods \cite{Khaneja_JMR_172_296305,Rice00,Brumer03} to engineer the control pulses producing the gates.

In this work we apply optimal control to find a fast and accurate pulse shape to speed up a controlled-Z gate between two qubits connected by a resonator \cite{Sillanpaa07,Majer07,Blais07,Plantenberg07,Liu07,Chow11,Dewes12,DiCarlo_nature_460_240,Fedorov12,Ghosh_PRA_87_022309}. The controlled-Z completes a universal gate set together with single-qubit rotations\cite{Vidal04}. The setting is motivated by superconducting qubits but has analogies in atomic physics \cite{Raimond01}, quantum dot \cite{Frey12,Petersson12}, and other resonator-based systems. We demonstrate the feasibility of these pulses by taking into account bandwidth limitations imposed by the experiment and provide a methodology for removing systematic errors that can practically affect the application of the control pulse.

The plan of the paper is as follows: In chapter \ref{sec:system} we describe the system setting as well as manually obtained methods to create CZ gates there. Sec. \ref{ch:grape} describes the implementation of optimal control to this system and its results and sec. \ref{ch:Errors} discusses potential error sources and their mitigation.

\section{System \label{sec:system}}
\subsection{Hamiltonian}

\begin{figure}[htbp!] \centering
 \includegraphics[width=0.48\textwidth]{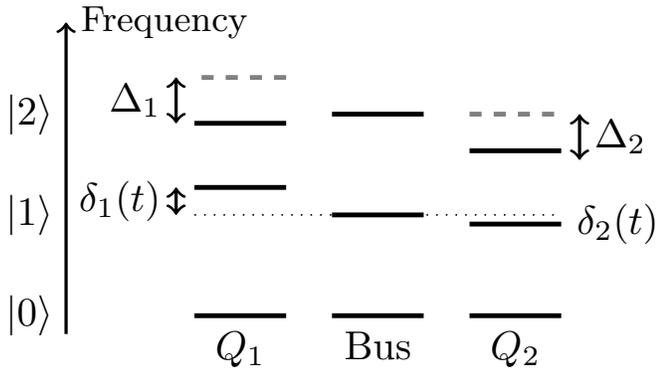}
 \caption{Sketch of the system where each element has three levels. The bus is harmonic and the qubits have anharmonicity $\Delta_k$ which can be dependent on the detunings in the case of the phase qubits. \label{Fig:LvlSketch}}
\end{figure}

In the following $\hbar=1$. The system of interest is made of two qubits coupled to a bus resonator, the qubits are sufficiently far apart so that their direct coupling can be neglected. They are modelled by three level non-linear oscillators. The third level accounts for leakage and in the case of the CZ gate can be populated to perform the gate. The individual qubit Hamiltonians are
\begin{equation}
 \hat H_{q_k}=\Delta_k\ket{2}_{k\,k\!\!}\bra{2}+\omega_k^{\phantom{+}}\hat\sigma_k^+\hat\sigma_k^-\, ,
\end{equation}
$\hat\sigma_k^+$ and $\hat\sigma_k^-$ respectively create and destroy one excitation in qubit $k$, $\hat{\sigma}_k^\pm =\sum_n |n\pm1\rangle_k\! _k\langle n|$. $\omega_k$ is the $0\leftrightarrow1$ transition frequency and $\Delta_k$ is the anharmonicity. The bus, with transition frequency $\omega_\text{b}$, is harmonic and posses three levels: $\hat{H}_b=\omega_\text{b}^{\phantom{\dagger}}\hat a^\dagger_\text{b}\hat a_\text{b}^{\phantom\dagger}$. The dipolar coupling strength between the bus and qubit $k$ satisfies $g_k\ll\omega_k$ and therefore the rotating wave approximation holds. The system's total Hamiltonian in this approximation is
\begin{align}
 \hat{H}=&~\omega_\text{b}^{\phantom{\dagger}}\hat{a}^\dagger_\text{b}\hat{a}_\text{b}^{\phantom{\dagger}}+\sum\limits_{k=1}^2 \Big[\Delta_k\ket{2}_{k\,k\!\!}\bra{2} +\omega_k^{\phantom{+}}\hat\sigma_k^+\hat\sigma_k^- \notag \\ +&~\frac{g_k}{2}\LR{\hat{\sigma}^+_k\hat{a}_\text{b}^{\phantom{\dagger}}+\hat{\sigma}^-_k\hat{a}^\dagger_\text{b}} \Big]\notag\,.
\end{align}
By the transformation $\hat{H}^R=R^\dagger\hat{H}R-i\dot{R}^\dagger R$ where
\begin{equation}
 \hat{R}=\LR{\sum\limits_{j=0}^2\exp\{-ij\omega_\text{b}t\}\ket{j}\bra{j}}^{\otimes3},
\end{equation}
we move to the rotating frame in which energies are measured with respect to the transition frequency of the bus. The Hamiltonian is 
\begin{equation}
 \hat{H}^R=\sum\limits_{k=1}^2\left[\delta_k(t)\hat{n}_{k}+\Delta_k\ket{2}_{k\,k\!\!}\bra{2} +\frac{g_k}{2}\LR{\hat{\sigma}^+_k\hat{a}_\text{b}^{\phantom{\dagger}}+\hat{\sigma}^-_k\hat{a}^\dagger_\text{b}}\right] \label{Eqn:H_rwa}
\end{equation}
where $\hat n_k=\hat\sigma_k^+\hat\sigma_k^-$ is the number operator for qubit $k$. The time dependence of the qubit-resonator detuning $\delta_k(t)=\omega_k(t)-\omega_\text{b}$ is made explicit to indicate the controls. The energy levels are sketched in Fig. \ref{Fig:LvlSketch}. The Hilbert space size is 27 dimensional and Hamiltonian (\ref{Eqn:H_rwa}) conserves the number of excitations. We make use of this to reduce the size of the Hilbert space since only computational states --- states in which the qubits have at most one excitation --- are of interest. Therefore we only need to study the 10 states with at most 2 quanta. The model is valid for transmons \cite{Koch_PRA_76_042319} and phase qubits \cite{Martinis02}. When dealing with the latter, the anharmonicity is a function of the qubit transition frequency and therefore a function of the controls $\Delta_k=\Delta_k(\delta_k(t))$. However for transmon qubits in the limit of large Josephson energy to charge energy it can be kept constant \cite{Koch_PRA_76_042319} to sufficient precision. 

\subsection{Analytic CZ Gate \label{Sec:Analytic_CZ}}
\begin{figure}[htbp!] \centering
 \includegraphics[width=0.48\textwidth]{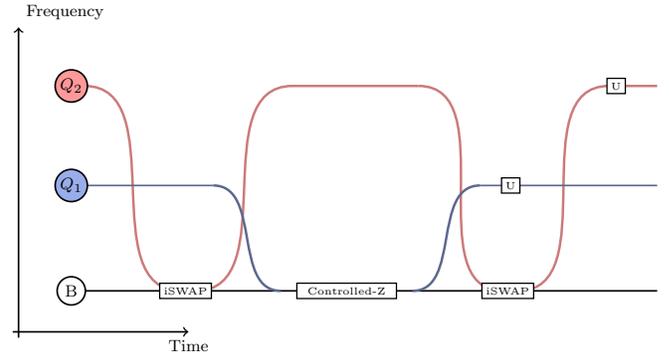}
 \caption{Sketch showing how the qubit's frequency is changed as function of time to create a CZ gate. \label{Fig:Analytic_CZ}}
\end{figure}

In the Qubit-Bus-Qubit system, the entangling gate needed to form a universal set of gates, is the CZ defined by $\ket{ij}\mapsto(-1)^{ij}\ket{ij}$. It is realized with 2 iSWAPs and a conditional rotation through a $\ket{2}$ state \cite{Mariantoni_Sci_334_6165}. A sketch of the pulse sequence is shown in Fig. \ref{Fig:Analytic_CZ}. The rotation through the $\ket{2}$ state only takes place when both qubits are in the $\ket{1}$ state; this can entangle the qubits. This $2\pi$ rotation is referred to as the Strauch gate \cite{Strauch_PRL_91_167005}. This sequence is slow due to the shifting of states in and out of the resonator. An improved analytic pulse sequence has been found in
Ref.  {\em et al.} \cite{Ghosh_PRA_87_022309}. This work considers an alternate approach based on numerical methods.

\subsection{Three Level Qubit and Bus \label{Sec:ThreeLvl}}
\begin{figure}[htbp!] \hfill
 \subfigure[\label{Fig:3LevelJC_E1}]{\includegraphics[width=0.48\textwidth]{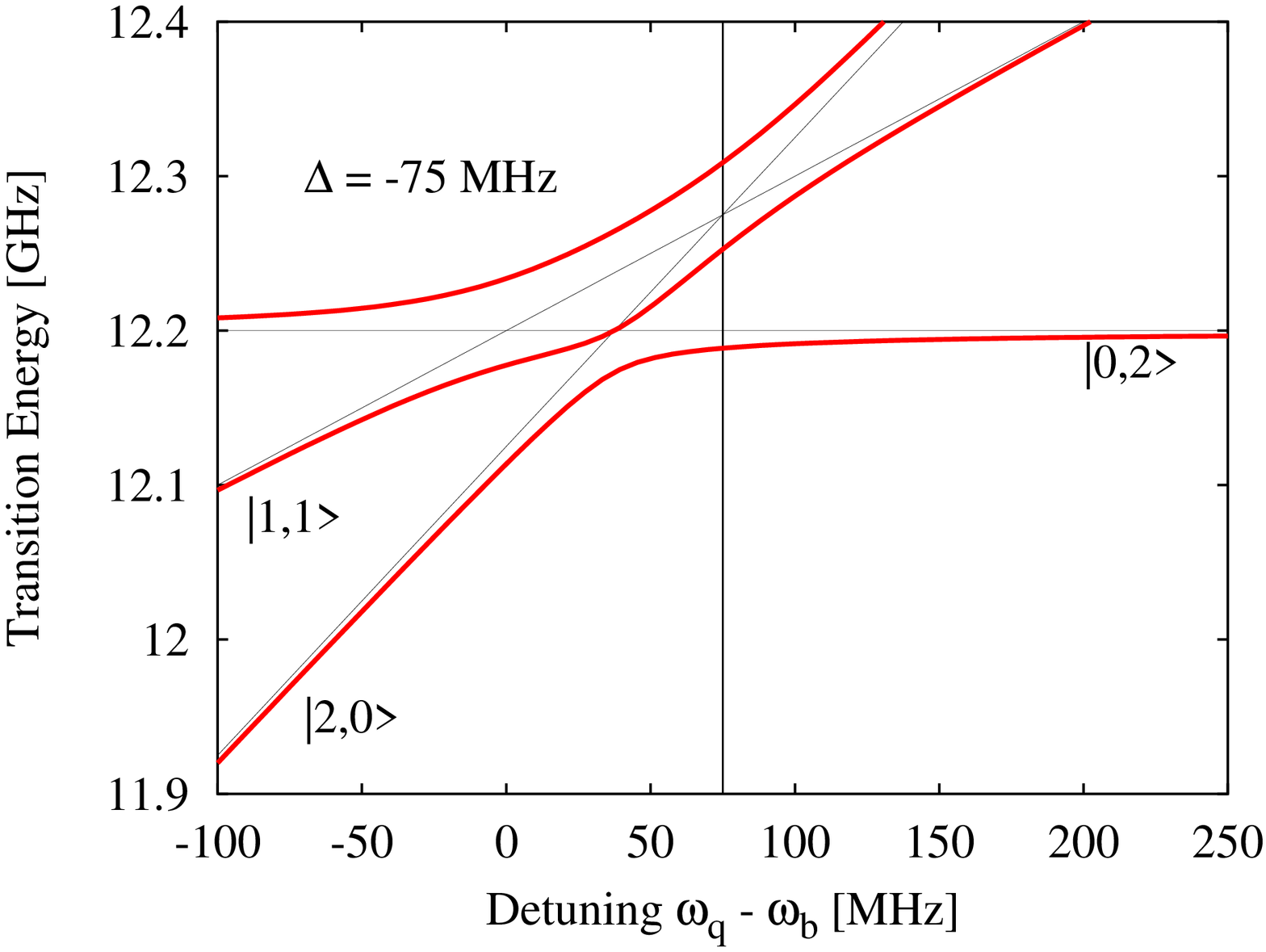}}  \hfill
 \subfigure[\label{Fig:3LevelJC_E2}]{\includegraphics[width=0.48\textwidth]{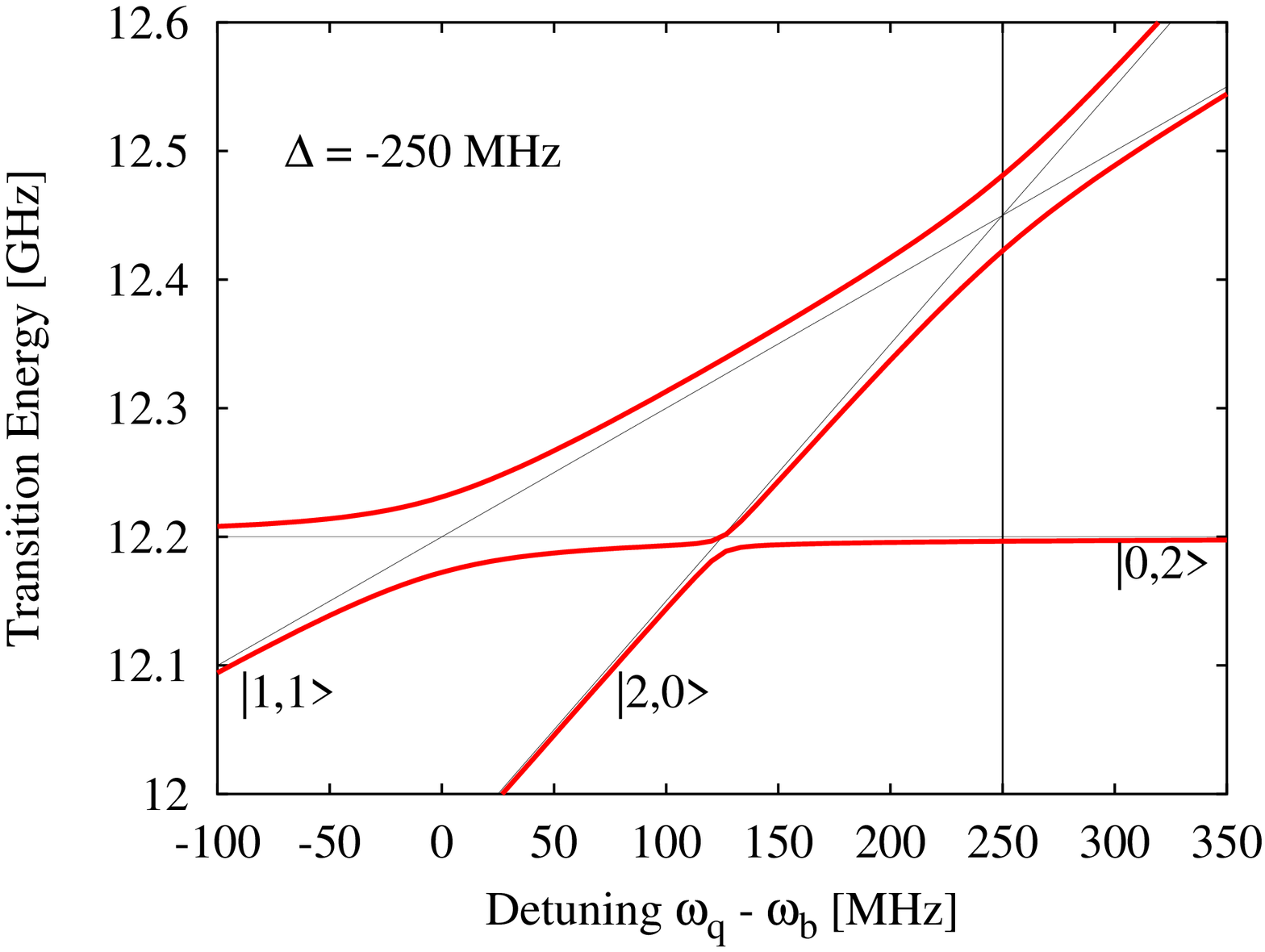}} \hfill
 \caption{Energy of the dressed states in the three level Jaynes-Cummings model with $\omega_\text{b}=6.1~\mathrm{GHz}$ and $T_\text{swap}=12~\mathrm{ns}$, i.e. the time it takes to swap a single excitation between the qubit and the bus. The three light black lines indicate the bare states whilst the vertical black line is $-\Delta/2$. \subref{Fig:3LevelJC_E1} Dressed states with $-75$ MHz anharmonicity and \subref{Fig:3LevelJC_E2} dressed states with $-250$ MHz anharmonicity. \label{Fig:3LevelJC_E}}
\end{figure}

\begin{figure}[htbp!] \hfill
 \subfigure[\label{Fig:3LevelJC_P1}]{\includegraphics[width=0.48\textwidth]{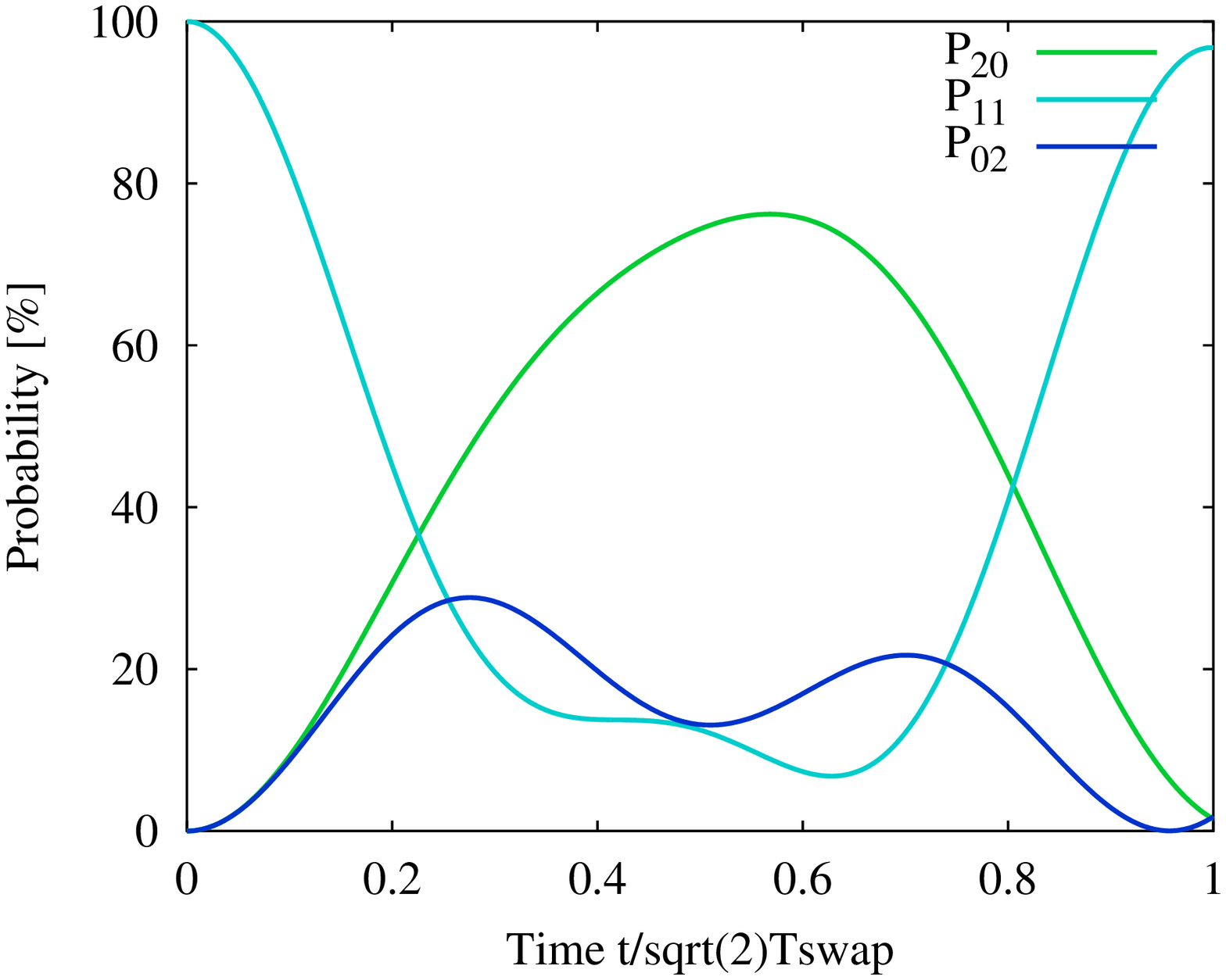}} \hfill
 \subfigure[\label{Fig:3LevelJC_P2}]{\includegraphics[width=0.48\textwidth]{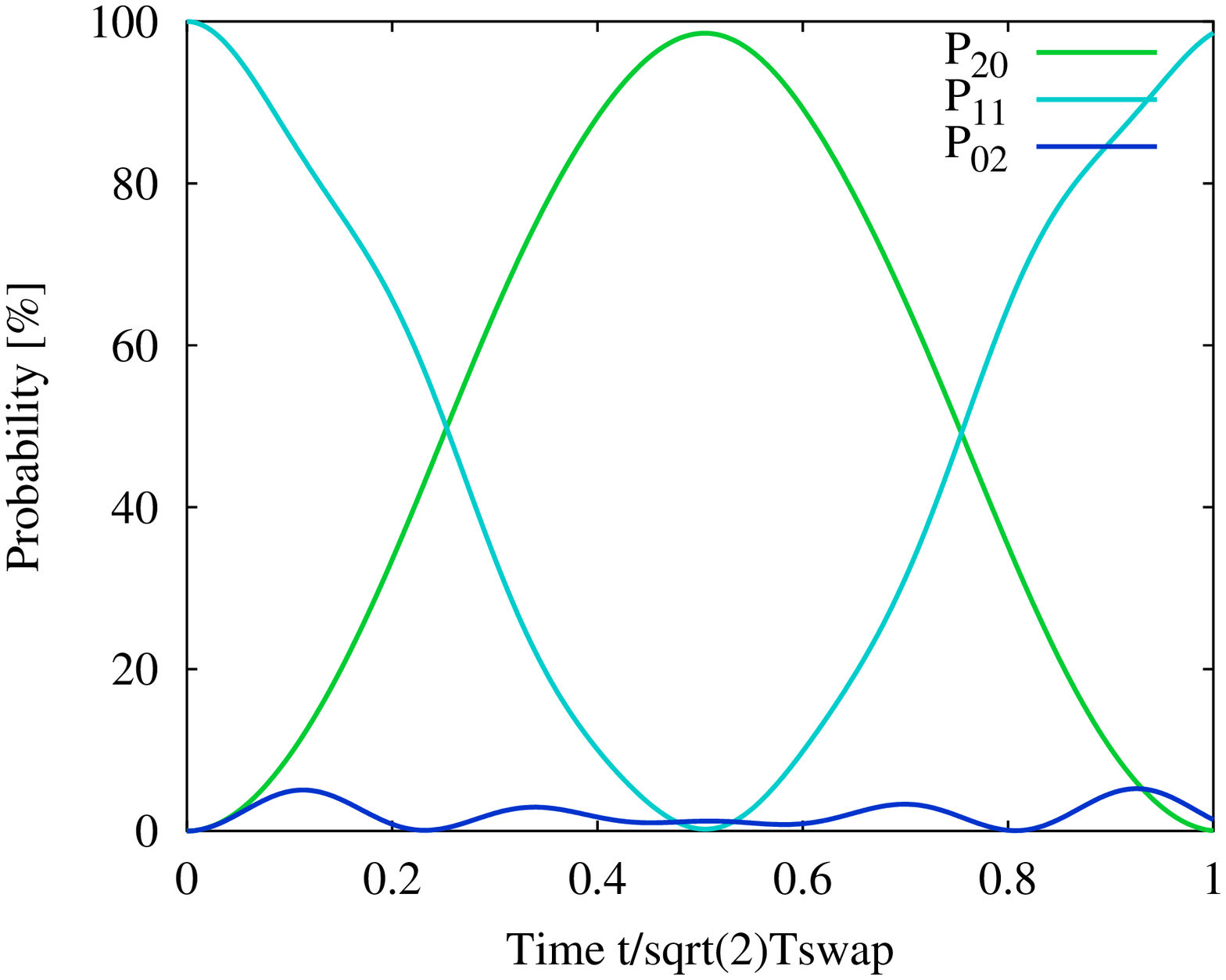}} \hfill
 \caption{Time evolution of the populations computed using the spectrum of Fig. \ref{Fig:3LevelJC_E} at $\delta=-\Delta$. Time is normalized to the duration of a $2\pi$ rotation through the qubit's $\ket{2}$ state. If the anharmonicity is too small the effect of the second state of the bus will be large. This degrades the fidelity of the Strauch gate. \subref{Fig:3LevelJC_P1} Population against time with $-75$ MHz anharmonicity and \subref{Fig:3LevelJC_P2} population against time with $-250$ MHz anharmonicity.\label{Fig:3LevelJC_P}}
\end{figure}

\begin{figure}[htbp!] \centering
 \includegraphics[width=0.48\textwidth]{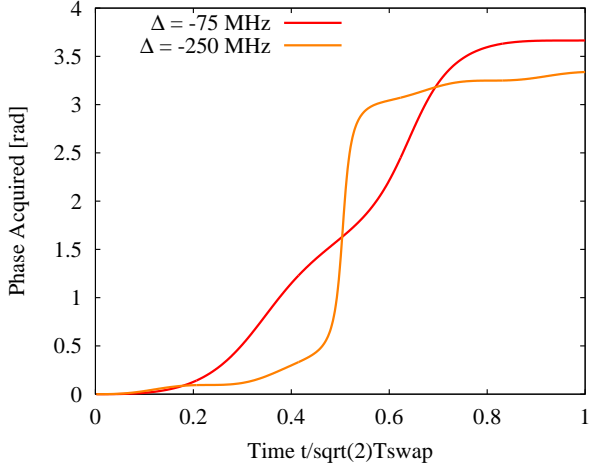}
 \caption{Phase difference between $\braket{1,1|\exp\{-i\hat Ht\}|1,1}$ and $\exp\{-i(2\omega_\text{b}-\Delta)t\}$. The discrepancy at the end of the gate is due to the presence of the unwanted $\ket{2}$ state of the bus. \label{Fig:3LevelJC_Phase}}
\end{figure}

The fidelity of a Strauch gate is degraded by the presence of other levels in the system. To illustrate this we consider a simplified model compared to Hamiltonian \eqref{Eqn:H_rwa}; a three level anharmonic qubit coupled to a resonator
\begin{align} \notag
 \hat{H}_\text{QB}=&~\underbrace{\omega_\text{b}\LR{\hat{a}^\dagger \hat{a}+\hat\sigma^+\hat\sigma^-}}_{\hat{H}_I} \notag \\ 
 +&~\underbrace{\delta \hat\sigma^+\hat\sigma^-+\Delta\Ket{2}\Bra{2}+\frac{g}{2}\LR{\hat{\sigma}^+\hat{a}+\hat{\sigma}^-\hat{a}^\dagger}}_{\hat{H}_{II}}
\end{align}
where $\delta=\omega_\text{q}-\omega_\text{b}$. This Hamiltonian conserves excitation number and thus is block diagonal with at most 3x3 blocks. For the block with $n$ excitations $\hat H_\text{I}$ is diagonal with identical values of $n\omega_\text{b}$. Focusing on $n=2$ the bare states are $\ket{2,0}$, $\ket{1,1}$ and $\ket{0,2}$. The Hamiltonian is
\begin{align}
\hat{H}^{(2)}=\begin{bmatrix} 2\omega_\text{b}+2\delta+\Delta & \sqrt{2}g & 0 \\ \sqrt{2}g & 2\omega_\text{b}+\delta & \sqrt{2}g \\ 0 & \sqrt{2}g & 2\omega_\text{b} \end{bmatrix} \label{Eqn:H_sub2}
\end{align}
An example of the eigenvalues of $\hat H^{(2)}$, for two different anharmonicities, are shown in Fig. \ref{Fig:3LevelJC_E}. The fine black lines represent energies of the uncoupled system, i.e. the bare states. A controlled-Z gate is made by a $2\pi$ rotation through the second excited state of the qubit i.e. $\ket{1,1}\circlearrowleft\ket{2,0}$. This is made possible by the anti-crossing indicated by the vertical black line in Fig. \ref{Fig:3LevelJC_E}. This is when the qubit's $\ket{1}\leftrightarrow\ket{2}$ transition is on resonance with the bus. Here the additional level $\ket{0,2}$ is an unwanted state; any population entering it will decrease the gate's fidelity. By inspection of Hamiltonian \eqref{Eqn:H_sub2}, the larger the anharmonicity is, the further away the $\ket{0,2}$ state is detuned. To clearly see it's influence, the time evolved population, shown in Fig. \ref{Fig:3LevelJC_P}, is computed with $\delta=-\Delta$ and for two different values of $\Delta$; one small and one large. When $\Delta$ is small, the Strauch gate performs badly as shown by Fig. \ref{Fig:3LevelJC_Phase}. With $-250$ MHz anharmonicity the leakage to $\ket{0,2}$ is at maximum 5\%, this is still large. Figure \ref{Fig:3LevelJC_Phase} shows that the phase difference at the end of the $2\pi$ rotation between the time evolved  state looping from and to $\ket{1,1}$ and the reference $\exp\{-i(2\omega_\text{b}-\Delta)t\}$ has a small deviation from $\pi$. The discrepancy is due to leakage to the $\ket{0,2}$ state. Such phenomena and multi-step swapping warranty a numerical approach to the problem of CZ gate design in the Qubit-Bus-Qubit architecture where the full Hamiltonian up to $n=2$ quanta is considered. Algorithms such as GRAPE and the quasi-Newton BFGS method \cite{Nocedal_Springer_2006} naturally suppress leakage since it decreases fidelity \cite{Rebentrost_PRB_79_060507}.

\section{Controlled-Z Gate Design by Gradient Ascent\label{ch:grape}}
Gradient ascent pulse shape engineering (GRAPE) numerically solves the problem of finding a control pulse that produces the desired time evolution operator within a given time \cite{Khaneja_JMR_172_296305}. In this work, the pulses are updated using the quasi-Newtonian BFGS method\cite{Nocedal_MC_35_773}. Hamiltonian (\ref{Eqn:H_rwa}) is separated into the drift and the two control control parts $\hat n_1$ and $\hat n_2$. This section describes the GRAPE implementation to the problem at hand covers the choice of fidelity function, the effect of the electronics and how to deal with non-linearities arising when using phase qubits. We then apply gradient ascent to systems with different parameter values to illustrate key features of the system. We also benchmark the numerical pulses on a system corresponding to real qubits.

\subsection{Fidelity Function}
We consider only unitary evolution. Thus, the overlap between the ideal $\hat U_\text{ideal}$ and achieved gate $\hat U$ serves as a fidelity function \cite{Khaneja_JMR_172_296305}
\begin{align}
 \Phi=\frac{1}{d^2}\left\vert\Trace{\hat U_\text{ideal}^\dagger \mathds{P}_Q\hat U\mathds{P}_Q}\right\vert^2 \label{Eqn:Phi}
\end{align}
$\mathds{P}_Q$ projects the time evolution operator onto the computational sub-space. Any leakage out of this sub-space will be detected as missing probability \cite{Rebentrost_PRB_79_060507}.

\subsection{Including Electronic Transfer Functions}
The arbitrary waveform generator (AWG) creating the control pulses has a limited bandwidth. Additionally the lines and remaining electronics between the AWG and qubits can distort the pulses. For this reason, the input control sent to the AWG will differ from the control applied by the qubits. In good approximation, this transfer is described by a linear causal transfer function \cite{Gustavsson_PRL_110_040502}. When optimizing the pulse shapes it must be ensured that the result is experimentally feasible. However the nature of the problem would require including numerical derivatives of measured transfer function data in the pulse optimization. We avoid this by convoluting the pulses with a Gaussian to suppress high frequencies
\begin{align}
 \delta_\text{qubit}(t)=\int\limits_{-\infty}^\infty\exp\left\{-\frac{(t-\tau)^2}{2\sigma^2}\right\}\delta(\tau)\mathrm{d}\tau
\end{align}
$\delta_\text{qubit}$ is the pulse shape that the qubit should see. The gradient is found with the chain rule \cite{Motzoi_PRA_84_022307}. The standard deviation $\sigma$ should be chosen to reflect the capabilities of the AWG. In an experimental implementation, it may be necessary to further optimize the pulse in the qubit control software to take into account the true transfer function, which must be measured due to its complex nature.

\subsection{Frequency Dependent Anharmonicity}
When optimizing pulses for a system where part of the Hamiltonian depends non-linearly on the controls the gradient rules of \cite{Khaneja_JMR_172_296305} must take the non-linearity into account. Such a situation can arise when optimizing pulses for phase qubits where the anharmonicity depends non-linearly on the qubit frequency. Appendix A of \cite{Machnes_PRA_84_022305} shows how to obtain the analytic formula of the gradient where the Hamiltonian depends linearly on the controls $\hat H(t)=\hat H_d+\sum_k \delta_k(t)\hat H_k$. Here is shown how to include non-linearities. We assume that in the Hamiltonian there are some parameters $\Delta_l$ that depend non-linearly on the controls, i.e. $\Delta_l=\Delta_l(\{\delta_k(t)\})$. The total Hamiltonian at time $j\Delta T$ becomes 
\begin{align}
\hat H(j\Delta T)=\hat H_d+\sum_k\delta_{kj}\hat H_k+\sum_l\Delta_l(\{\delta_{kj}\})\hat H_{l,\text{nl}}\, .
\end{align}
The gradient of the time evolution operator $\hat U_j$ of time-slice $j$, with respect to pixel $\delta_{kj}$ of control $k$, is only sensitive to small variations around the value assumed by that pixel. Therefore we linearise the Hamiltonian at each iteration. If pixel $kj$ assumes the value $\delta_{kj}^{(n)}$ at iteration $n$ the Hamiltonian can be approximated by $\hat H(j\Delta T)\simeq \hat H'_d+\sum_k\delta_{kj}\hat H'_k$ where the drift and controls of this linearised Hamiltonian are
\begin{align}
  \hat H'_d=&~\hat H_d+\sum_l\Delta_l\left(\left\{\delta_{kj}^{(n)}\right\}\right)\hat H_{l,\text{nl}}\, , \\
  \hat H'_k=&~\hat H_k+\sum_l\left.\frac{\partial\Delta\LR{\{\delta_{kj}\}}}{\partial\delta_{kj}}\right|_{\delta_{kj}^{(n)}}\hat H_{l,\text{nl}}\, .
\end{align}
This enables us to compute the gradient using the rules given in \cite{Khaneja_JMR_172_296305,Machnes_PRA_84_022305}. At each iteration the control fields change and so do the linearized Hamiltonians $\hat H'_d$ and $\hat H'_k$.

In the case of phase qubits, the dependency of the anharmonicity $\Delta_k$ on the qubit frequency $\delta_k$ can either be found numerically with a discrete variable representation \cite{Colbert_JCP_96_19821991} of the qubit's full potential or through measurement with high-power spectroscopy \cite{Yamamoto_PRB_82_184515}.

\subsection{Numerical Results\label{ch:toy_results}}

Here we assume that both qubits have the same anharmonicity $\Delta_1=\Delta_2=\Delta=-0.1\omega_\text{b}$ and coupling $g_1=g_2=g=0.02\omega_\text{b}$. Time will be indicated in dimensionless units of $tg$ with $g$ in radians $s^{-1}$. The desired gate fidelity is $1-10^{-4}$. At the start and end of the gate both qubits are far of resonance at their parking frequencies. This is included in the code by adding several buffer pixels, held at a constant detuning, before and after the gate.

\begin{figure}
 \subfigure[\label{Fig:T450_Pulse_Pulse}]{\includegraphics[width=0.48\textwidth]{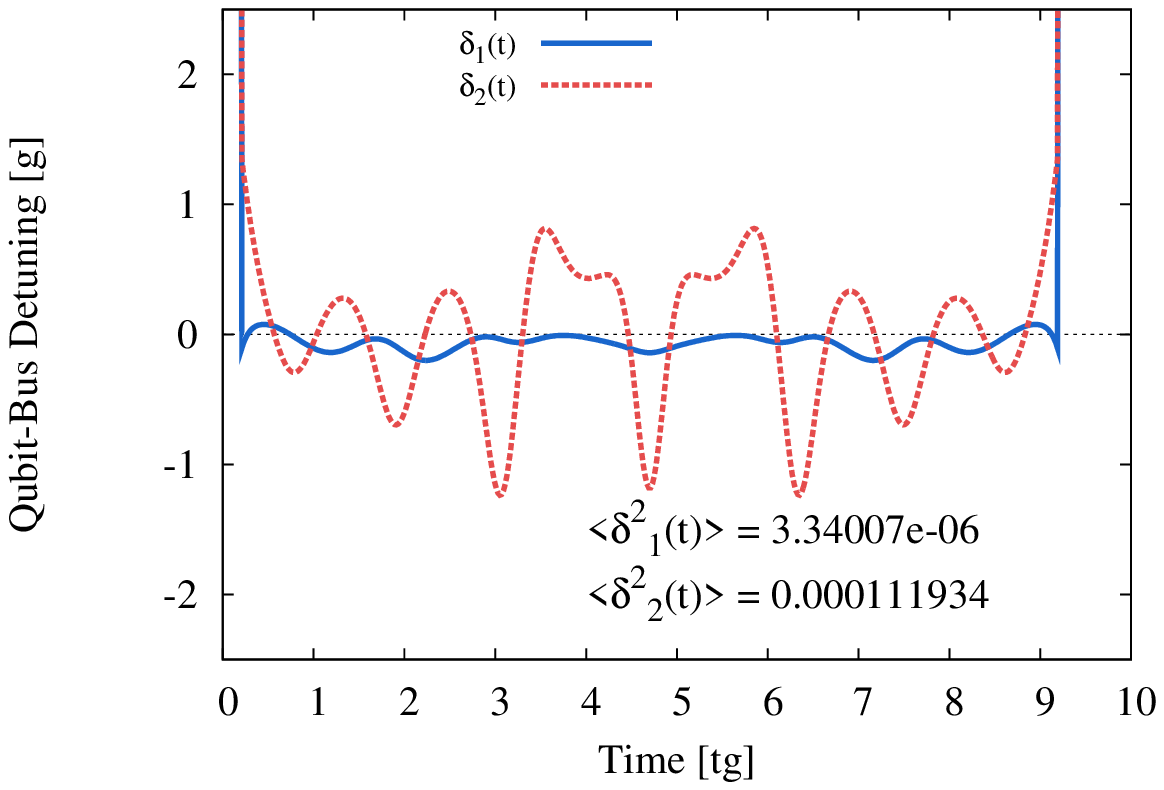}}
 \subfigure[\label{Fig:T450_Pulse_DFT}  ]{\includegraphics[width=0.48\textwidth]{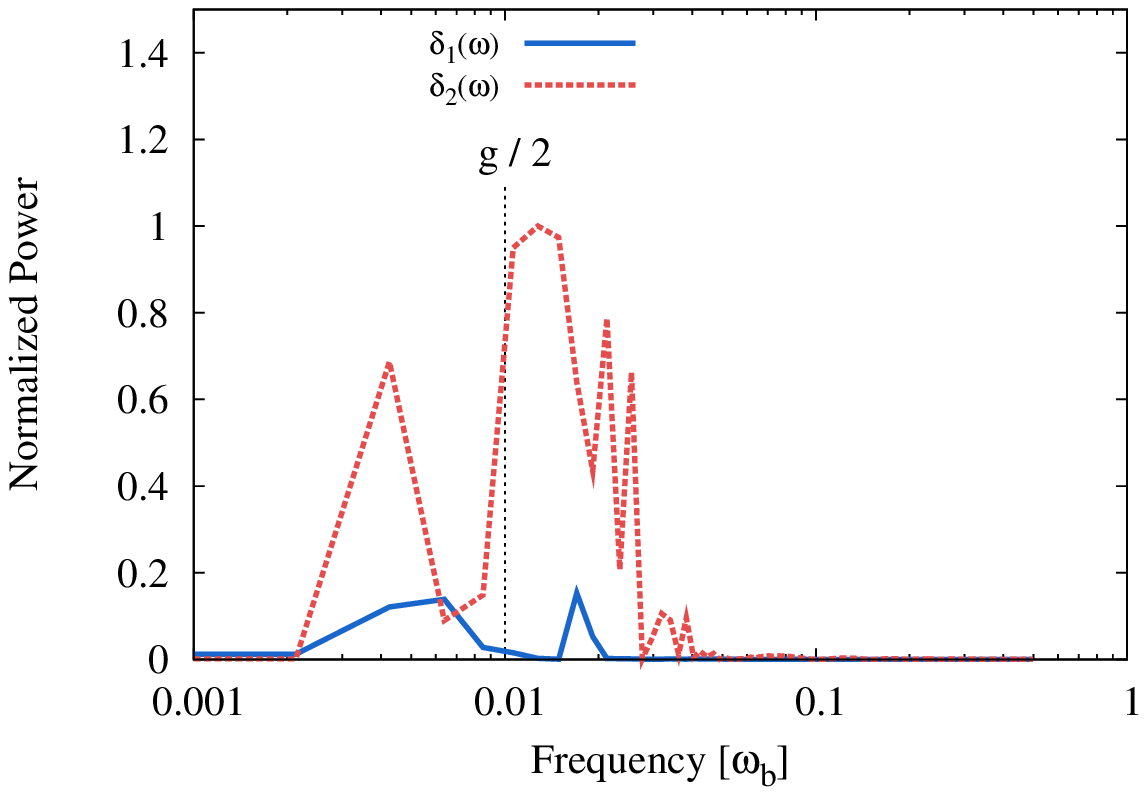}}
 \caption{\subref{Fig:T450_Pulse_Pulse} Control pulses for a gate time slightly above $T_\text{QSL}$. \subref{Fig:T450_Pulse_DFT} Discrete Fourier Transform of \subref{Fig:T450_Pulse_Pulse} showing that most of the spectral power is at small frequencies of order $g$. \label{Fig:T450_Pulse} }
 \end{figure}

The control pulses, found without Gaussian convolution, for a gate time of $9~tg$ are shown in Fig. \ref{Fig:T450_Pulse_Pulse}. Figure \ref{Fig:T450_Pulse_DFT} shows the Discrete Fourier Transform (DFT) of these pulses: most of the oscillations in $\delta_1(t)$ and $\delta_2(t)$ have frequency components of the order of the qubit-bus coupling constant $g/\sqrt{2}$. This is because the CZ gate is made using $\ket{2}$ states.

Figure \ref{Fig:T450_Pulse} shows that the controls for qubit two oscillate at much larger amplitude than those for qubit one. We will later demonstrate that qubit 2 and the resonator populate their $\ket{2}$ states similar to the Strauch gate in the pulse sequence. As the CZ gate is symmetric under the exchange of qubits, a control-target terminology to distinguish these qubits would be inappropriate. Instead,
 the qubit with smaller oscillations will be referred to as \emph{Ginger} whereas the other will be called \emph{Fred}. The next section explores what motivates the
symmetry breaking apparent in these pulses

\subsubsection{Effect of Anharmonicity \label{Sec:GRAPE_anharmon}}
\begin{figure}
 \subfigure[\label{Fig:AnH_D03_D01}]{\includegraphics[width=0.48\textwidth]{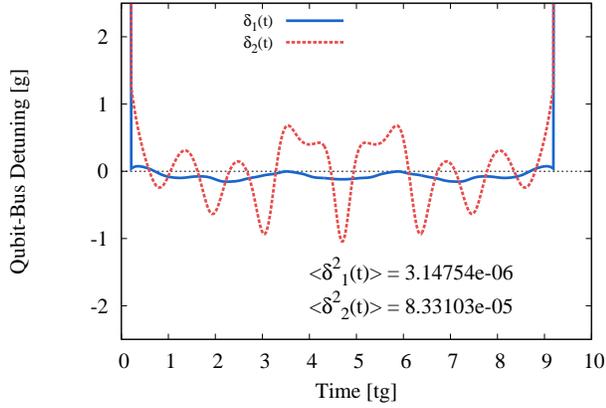}}
 \subfigure[\label{Fig:AnH_D01_D03}]{\includegraphics[width=0.48\textwidth]{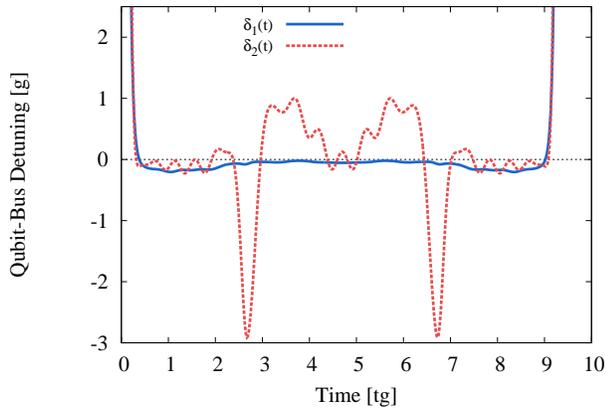}}
\caption{Comparison of the choice of Ginger and Fred. \subref{Fig:AnH_D03_D01} $(\Delta_1,\Delta_2)=(-0.3,-0.1)$. Qubit 2, chosen as Fred, is the most linear. The optimization was successful reaching $\Phi=99.99\%$. \subref{Fig:AnH_D01_D03} $(\Delta_1,\Delta_2)=(-0.1,-0.3)$. Qubit 2, chosen as Ginger, is the most linear. The optimization was unsuccessful reaching only $99.95\%$. In both cases asymmetric initial conditions were used to force GRAPE to chose qubit 2 as Fred. Only the case shown in figure (a) resulted in good convergence.}
\end{figure}

As just stated, the CZ gate is symmetric under qubit exchange, however, the underlying Hamiltonian need not be. The main aspect breaking the symmetry is the anharmonicity of the qubit. This was studied with several different combinations of qubit anharmonicities: $\Delta_1,\Delta_2\in\{-0.1,-0.2,-0.3\}$. The allowed gate time was $9~tg$. Because the Strauch method uses the $\ket{2}$ state of the bus, the more linear qubit takes on the role of Fred since it is easier to drive the  $\ket{1,1}\leftrightarrow\ket{0,2}$ transition with the bus.The reason is:  the greater the anharmonicity, the greater the qubit has to move away from the $\delta=0$ qubit-bus resonance, which is also crucial for $\ket{0,1}\leftrightarrow\ket{1,0}$ exchanges. Figures \ref{Fig:AnH_D03_D01} and \ref{Fig:AnH_D01_D03} show two pulses for which $(\Delta_1,\Delta_2)=(-0.3,-0.1)$ and $(\Delta_1,\Delta_2)=(-0.1,-0.3)$ respectively. In both cases the search was nudged by means of asymmetric initial conditions, to chose qubit two as Fred. In the first case when the most linear qubit was chosen as Fred the target gate error of $10^{-4}$ was reached. When the wrong qubit was assigned as Fred in the initial conditions, the code was not able to reach the target gate fidelity reaching only $1-\Phi\simeq5.5\cdot10^{-3}$. The choice as to which qubit assumes which role can either be enforced through asymmetric initial conditions or left up to GRAPE/BFGS with symmetric initial conditions. In the latter case the algorithm converges slower in the initial steps before numerical approximations break the symmetry. The anharmonicity of Ginger does not play such an important role as Fred's. Figure \ref{Fig:T450_Pulse} shows a pulse with $(\Delta_1,\Delta_2)=(-0.1,-0.1)$. The pulse of Fred is almost identical to the one in Fig. \ref{Fig:AnH_D03_D01}. However since Ginger is more linear than in Fig. \ref{Fig:AnH_D03_D01} its control pulse has picked up some additional modulation which could be to minimize leakage to the qubit $\ket{2}$ state.

\begin{figure}[htbp!] \centering
 \includegraphics[width=0.48\textwidth]{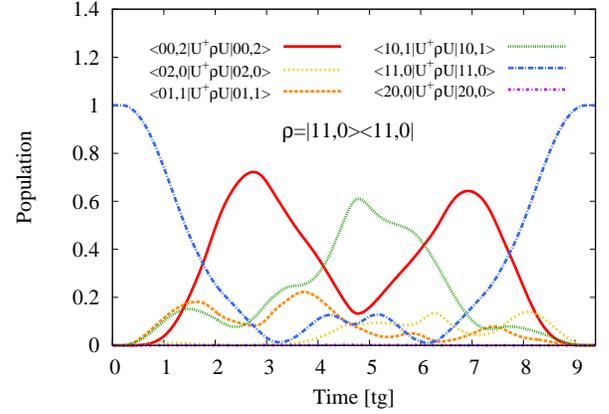}
 \caption{Populations assuming the input state is $\rho_{in}=\ket{11,0}\bra{11,0}$ for the pulse of Fig. \ref{Fig:AnH_D03_D01}. It shows that the $\ket{2}$ state of the bus is highly solicited to realize the CZ gate. However the $\ket{2}$ state of the most non-linear qubit is not used at all. \label{Fig:AnHPop_D03_D01}}
\end{figure}

The populations associated to the pulse shown in Fig. \ref{Fig:AnH_D03_D01} are displayed in Fig. \ref{Fig:AnHPop_D03_D01}; the $\ket{2}$ state of the bus is highly used. Some of the excitation is transferred to the $\ket{2}$ state of Fred but the $\ket{2}$ state of Ginger remains empty, confirming our interpretation of the role of both qubits.

\begin{figure}[htbp!] \centering
 \includegraphics[width=0.48\textwidth]{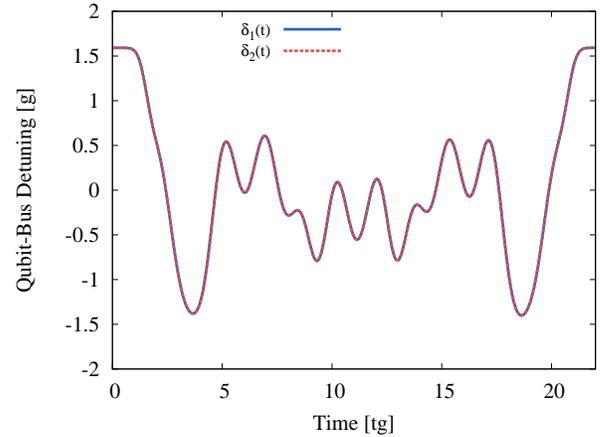}
 \caption{Control pulse with perfect fidelity (up to machine precision) for identical qubits and symmetric initial conditions. The control pulses producing the CZ gate are identical for both qubits showing that the two pulses need not be asymmetric. The qubit parameters were $g_1/2\pi=g_2/2\pi=50$ MHz and $\Delta_1=\Delta_2=60$ MHz. \label{Fig:SymmetricPulse}}
\end{figure}

However the pulses need not be asymetric. If both qubits are identical and the initial conditions are symmetric, the resulting pulse sequence will be symmetric. Such a symmetric pulse is shown in Fig. \ref{Fig:SymmetricPulse}.

\subsubsection{Influence of impulse response}
\begin{figure}[htbp!] \centering
 \includegraphics[width=0.48\textwidth]{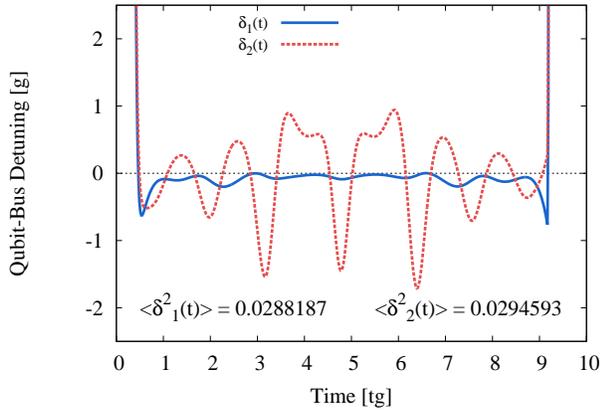}
 \caption{Effect of the filter function on the pulse sequence. The pulse from Fig. \ref{Fig:T450_Pulse} was used as a starting point for the gradient ascent. Given that most of the spectral weight was initially at low frequencies, the gaussian convolution has hardly any effect and the target fidelity of $99.99\%$ is retrieved after only a few iterations. \label{Fig:GausFiltPulse}}
\end{figure}

As can be seen from Fig. \ref{Fig:T450_Pulse_DFT} the DFT of the unfiltered pulse has almost all its power at low frequencies. This suggests that introducing a filter function in GRAPE should not significantly deteriorate the gate's performance. Therefore, in the control landscape, the optimal solutions with and without filter function should lie close together. The procedure is first to search for a pulse without the filter function and then to rerun the optimization with the filter function using the previously found pulse as the initial condition. The Gaussian impulse response has standard deviation of $\sigma\cong4~\omega^{-1}_b$, chosen so that the 3 dB attenuation lies slightly above $g=0.02\omega_\text{b}$. This function was then used to find a pulse sequence with the pulses shown in Fig. \ref{Fig:T450_Pulse} as starting point. The output is shown in Fig. \ref{Fig:GausFiltPulse}. As seen from the figures, the pulse found with the filter function is almost identical to the one found with a perfect impulse response. However the sharp corners have been smoothed out due to the high frequency filtering. This is encouraging since typical AWGs have a bandwidth of 500 MHz and most coupling strengths are in the range $20-70~\mathrm{MHz}.$ Given the small effect of the impulse response, the subsequent optimization will be done in one step using Gaussian convolution.

\subsection{Benchmarking \label{ch:phase_results}}
To benchmark the performance of the numerical pulses against existing pulses, the GRAPE method is applied to phase qubits in the RezQu architecture. The values\footnote{The values for the parameters in the Hamiltonian correspond to a sample of the John Martinis group.} for the parameters in the Hamiltonian are shown in Tab. \ref{Tab:JM_Values}. The non-linear behaviour of the qubit's anharmonicity was determined by high power spectroscopy \cite{Yamamoto_PRB_82_184515}. The anharmonicities for this chip are very low and as indicated from section \ref{Sec:ThreeLvl}, would produce Strauch gates with extremely low fidelities.

\begin{table}[htbp!] \centering
 \caption{Parameters of the phase qubits. These values were used in the pulses presented in this document. The swap bus time $T_\text{swap}$ is the time required to swap a quanta between the qubit and bus, i.e. $\ket{1,0}\rightarrow\ket{0,1}$. It is related to coupling strength by $g_k=(2T_{\text{swap},k})^{-1}$. \label{Tab:JM_Values}}
   \begin{tabular}{l l l r l} \hline\hline
    Element & \multicolumn{2}{l}{Parameter} & Value & unit \\\hline
    Bus & $\omega_\text{b}$ & Frequency & 6.1 & GHz \\
    Qubit 1 & $\omega_1$ & parking frequency \hspace{0.5cm} & 6.778 & GHz \\
    & $\Delta_1$ & Anharmonicity & -71 & MHz \\
    & & swap bus time & 12.6 & ns \\
    & $g_1$ & coupling strength & 40 & MHz \\
    Qubit 2 & $\omega_2$ & parking frequency & 6.607 & GHz \\
    & $\Delta_2$ & Anahamonicity & -59 & MHz \\
    & & swap bus time & 9.2 & ns \\
    & $g_2$ & coupling strength & 54 & MHz \\ \hline\hline
   \end{tabular}
\end{table}

\begin{figure}[htbp!] \centering
  \includegraphics[width=0.48\textwidth]{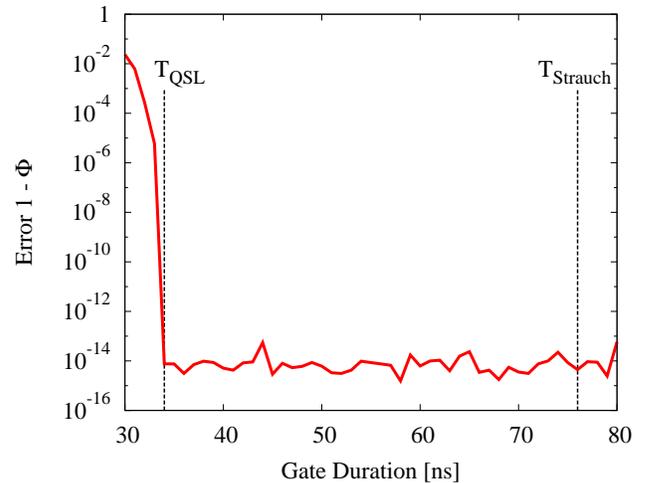}
  \caption{Scan of the gate duration to find the quantum speed limit for phase qubits with the values of Tab. \ref{Tab:JM_Values}. The found quantum speed limit $T_\text{QSL}=34~\mathrm{ns}$ is twice as fast as the sequential pulse using the Strauch gate which takes $T_\text{Strauch}=76~\mathrm{ns}$. Above the quantum speed limit, the numerical pulses are perfect up to machine precision. \label{Fig:JM_Tcrit}}
\end{figure}

\begin{figure}[htbp!] \centering \hfill
 \subfigure[\label{Fig:JM_Tcrit_Pulse}]{\includegraphics[width=0.48\textwidth]{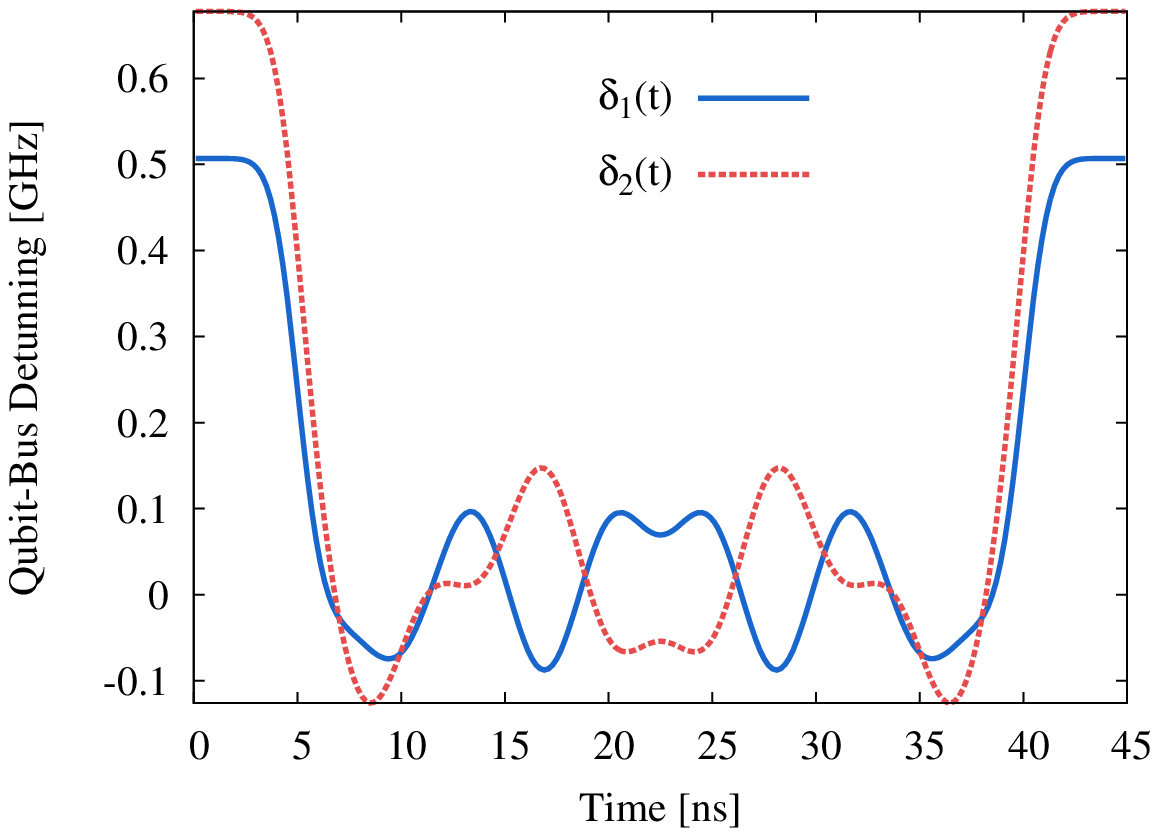}} \hfill
 \subfigure[\label{Fig:JM_Tcrit_Pulse_FT}]{\includegraphics[width=0.48\textwidth]{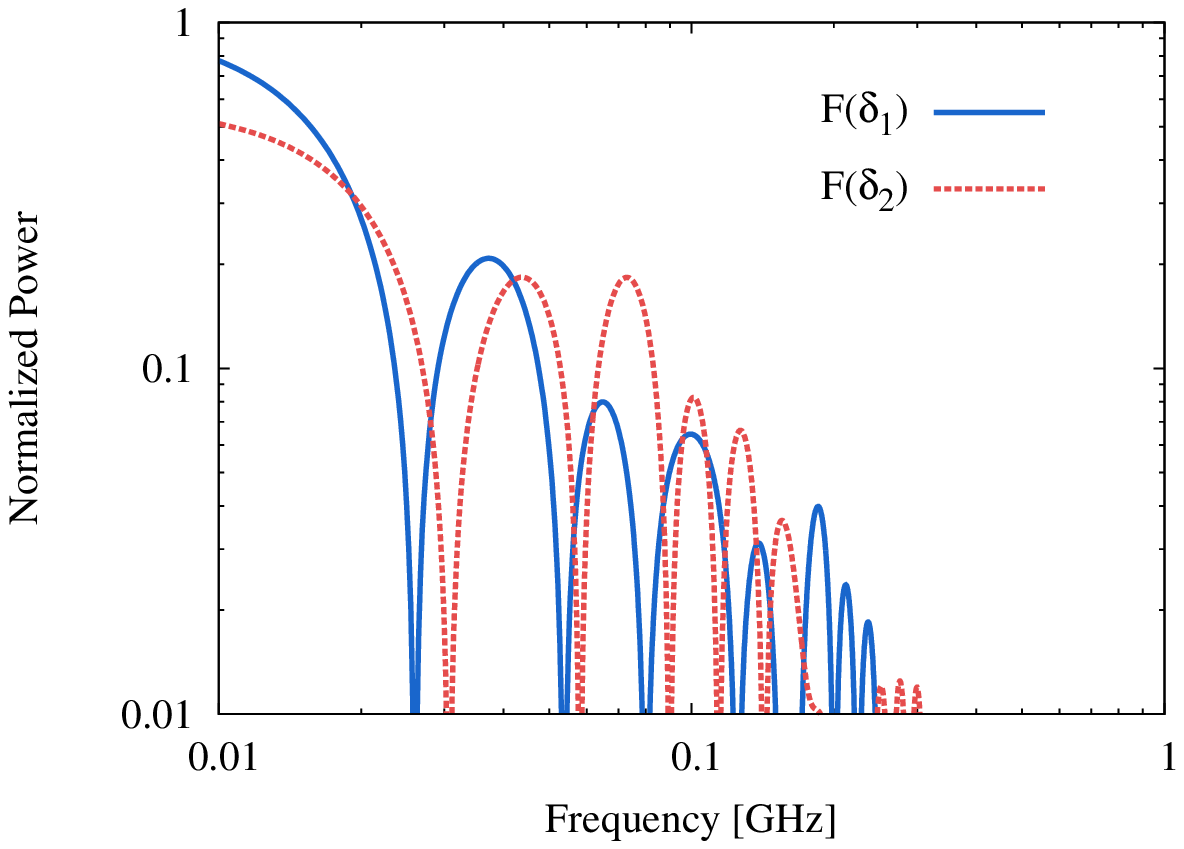}} \hfill
 \caption{Summary of a numerical CZ gate design. Despite the low anharmonicities the pulses are able to reach very high fidelities. \subref{Fig:JM_Tcrit_Pulse} Control pulse with $1-10^{-14}$ intrinsic fidelity. The gate time is slightly above the quantum speed limit, i.e. $T_\text{Gate}=35~\mathrm{ns}$. \subref{Fig:JM_Tcrit_Pulse_FT} Normalized spectrum of the ZPA corresponding to the GRAPE pulse shown in \subref{Fig:JM_Tcrit_Pulse}. Almost all the spectral power is within a few hundred MHz, thus the pulse is experimentally feasible.}
\end{figure}

\begin{figure}[htbp!] \centering
 \includegraphics[width=0.48\textwidth]{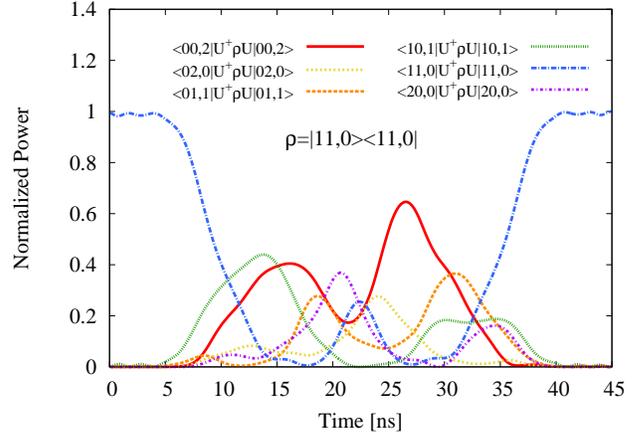}
 \caption{$~$Populations of the two excitation sub-space associated to the GRAPE pulse shown in Fig. \ref{Fig:JM_Tcrit_Pulse}. As can be seen, an analytical picture similar to that shown in section \ref{Sec:Analytic_CZ} is no longer possible. \label{Fig:JM_Tcrit_Pulse_Fpop}}
\end{figure}

In some situations the time it takes for a given state to evolve into an orthogonal state is bounded from below. This lower bound is the quantum speed limit (QSL) \cite{Schulman_LNP_734_107,Vaidman_AJP_60_182}. This sets a minimum time $T_\text{QSL}$ in which a gate can be done. When the gate time is above this speed limit, numerical pulses have perfect fidelity up to machine precision. This is shown in Fig. \ref{Fig:JM_Tcrit} where the gate time is progressively decreased. As long as $T_\text{Gate}>T_\text{QSL}$ the pulse error is zero down to machine precision. For the system with values given by \ref{Tab:JM_Values}, the QSL is less than half of the time it takes the analytic pulse sequence of Fig. \ref{Fig:Analytic_CZ}. We find $T_\text{QSL}=34~\mathrm{ns}$. Below the quantum speed limit the fidelity degrades very rapidly. A machine precision errorless pulse is shown in Fig. \ref{Fig:JM_Tcrit_Pulse}. As seen in Fig. \ref{Fig:JM_Tcrit_Pulse_FT}, a DFT shows that there is hardly any spectral power above $500~\mathrm{MHz}$ thus making the pulse experimentally realistic. Fig. \ref{Fig:JM_Tcrit_Pulse_Fpop} shows the populations in the two-excitation subspace illustrating the increased complexity of these fast gates, defying, for now, a simple physical picture.

\section{Error sources and mitigation strategies \label{ch:Errors}}
The previous section showed that CZ gates with arbitrary intrinsic fidelities can be generated even for low anharmonicity qubits. However, in experimental conditions these high quality pulses are rapidly degraded by various errors. The following section reviews them and discusses how to overcome them.

\subsection{Intrinsic Pulse Robustness}
\begin{figure}[htbp!]\centering
 \includegraphics[width=0.48\textwidth]{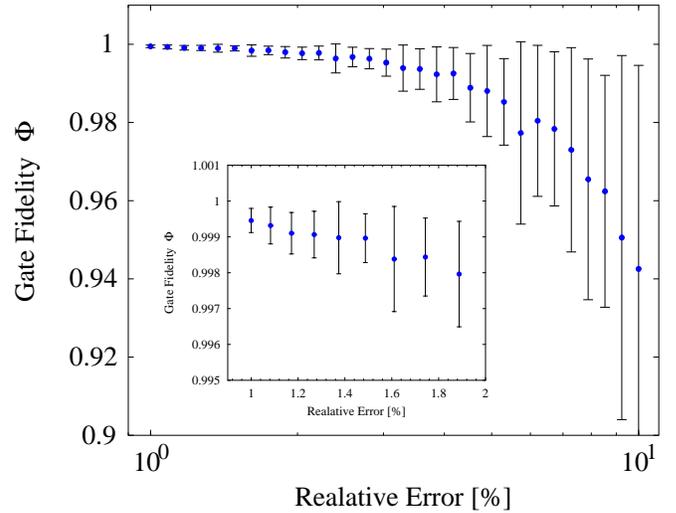}
 \caption{Error introduced by randomly changing the amplitude of the controls. The relative error is expressed in $\Delta \delta_{k_j}/\delta_{k_j}$. The system considered was the Qubit-Bus-Qubit with constant anharmonicity. \label{Fig:PulseRobustness}}
\end{figure}

Gradient ascent engineered pulses enjoy an almost null first derivative with respect to the individual control pixels. Thus to first order, random fluctuations of the pulse amplitude does not severely impact the fidelity. This was checked by perturbing the controls with white Gaussian noise with a standard deviation given by $\sigma_E=\Delta \delta_{k_j}/\delta_{k_j}$. Figure \ref{Fig:PulseRobustness} shows that a 1\% relative variation of the control field amplitude decreases a $99.99\%$ intrinsic fidelity pulse to $99.95\%$. Therefore random fluctuations in pulse amplitude are of little consequence on these pulses \cite{Spoerl07,Montangero07}. 

\subsection{Systematic Errors}
Some systematic errors will effect the pulses in a more significant way than the random fluctuations of pulse amplitude. There are three main errors that have been identified: calibration errors, timing errors and parameter errors. All are described below.

\subsubsection{Calibration Errors: Control Pulse Amplitude to Qubit Frequency \label{Sec:Err_Calibration}}
\begin{figure}[htbp!] \centering
 \includegraphics[width=0.48\textwidth]{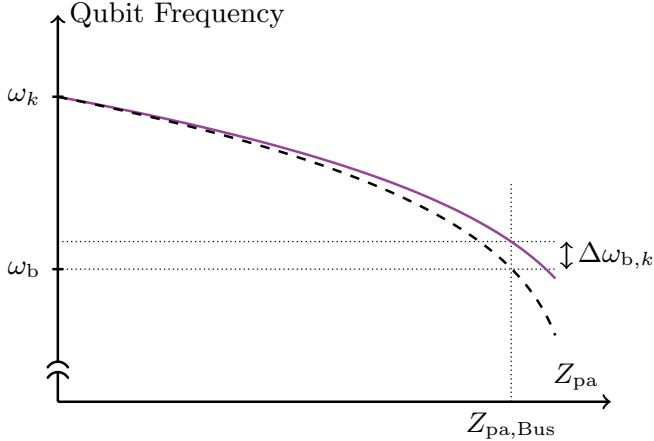}
 \caption{Calibration errors result in a DC offset of the pulse. The ``true'' calibration curve -- dashed line -- is approximated by the solid line which is the measured calibration curve. This discrepancy causes the qubit-bus resonance to be missed by $\Delta\omega_{\text{b},k}$. \label{Fig:CalErr}}
\end{figure}

\begin{figure}[htbp!] \centering
 \includegraphics[width=0.48\textwidth]{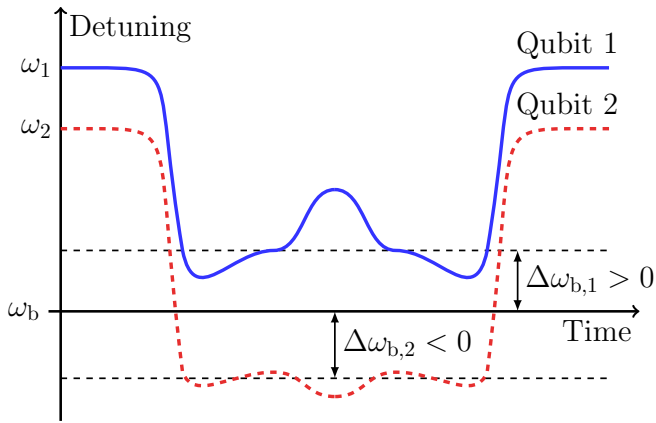}
 \caption{$~$DC offset in the pulse amplitudes. The qubit parking frequency $\omega_{Q_k}$ is left unchanged. However, the resonance point is missed; the pulses perform their oscillations around $\omega_\text{b}+\Delta\omega_{\text{b},k}$ instead of $\omega_\text{b}$. \label{Fig:BusErrSketch}}
\end{figure}

Although we optimize the qubit frequency in our numerics, the true control is the amplitude of the Z pulse (ZPA), a voltage pulse applied to the qubit. The ZPA is related to the frequency of the qubit through a calibration curve. This curve must be measured using spectroscopy and errors in it will cause errors in qubit frequency, see Fig. \ref{Fig:CalErr} for a sketch. The bus frequency does not enter the Hamiltonian \eqref{Eqn:H_rwa}. However it must be known so as to give the qubits the right ZPA to put them on resonance with the bus. A constant and systematic shift of the pulse with respect to the resonance point produces phase and leakage errors. The situation is sketched in Fig. \ref{Fig:BusErrSketch} where qubit 1 undershoots the bus by an amount $\Delta\omega_{\text{b},1}$ and qubit 2 overshoots the bus by $\Delta\omega_{\text{b},2}$.

\begin{figure}[htbp!] \centering
 \includegraphics[width=0.48\textwidth]{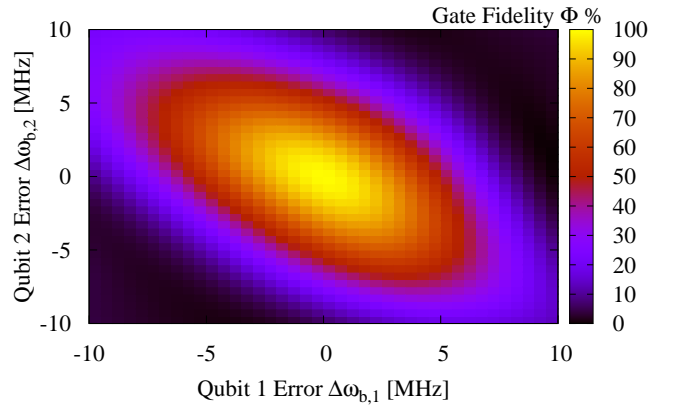}
 \caption{Loss of fidelity due to over and undershoot of the bus-qubit resonance frequency arising from systematic calibration errors. As can be seen errors on the bus frequency of less than 0.1\% ruin the pulses. \label{Fig:PulseNonRobustFidel}}
\end{figure}

Off resonance from the bus, calibration errors have little effect since qubit and resonator cannot exchange quanta. Therefore this error is modeled by a systematic shift in the qubit frequency changing the resonance point with the bus
\begin{equation}
 \delta_k(t)\mapsto\delta_k(t)+\Delta\omega_{\text{b},k}.
\end{equation}
This shift also displaces the qubit parking frequency, which, in experiment, is typically held constant at all times \cite{Galiautdninov12}. This discrepancy between experiment and model is acceptable due to the lack of exchange of quanta far of resonance. The Hamiltonian with error terms is
\begin{align}
 &\hat{H}^R=\underbrace{\sum\limits_{k=1}^2\delta_k(t)\hat n_k}_{\text{Controls}}   \label{Eqn:H_rwa_Err} \\ \notag
 &+\underbrace{\sum\limits_{k=1}^2\left[\Delta\omega_{\text{b},k}\hat{n}_k+\Delta_k\hat{\Pi}_{2,k}
 +\frac{g_k}{2}\LR{\hat{\sigma}^+_k\hat{a}_\text{b}^{\phantom{\dagger}}+\hat{\sigma}^-_k\hat{a}^\dagger_\text{b}}\right]}_{\text{Drift with errors}}\, .
\end{align}
The effect of the calibration error on the Fidelity (\ref{Eqn:Phi}) is shown in Fig. \ref{Fig:PulseNonRobustFidel}. A pulse was first optimized with  $(\Delta\omega_{\text{b},1},\Delta\omega_{\text{b},2})=(0,0)$ and then the fidelity is recomputed for different values of the error. The central high fidelity region is very small; although the first derivative $\partial\Phi/\partial \Delta\omega_{\text{b},k}$ is close to zero near the optimum the second derivative is very strong. This shows how small errors ruin the pulse.

If a single control amplitude at a given time is viewed as a degree of freedom, pulse optimization is a highly under-constrained non-linear problem given the limited number of independent parameters in the target gate. Robust control exploits the surplus of degrees of freedom to make a pulse sequence robust over a larger parameter range \cite{Khani_PRA_85_022306}. However in this case such methods fail since the error is on the control Hamiltonians and not the drift. To remove the calibration error a different approach must be used.

We propose to manually introduce a controllable DC offset in the pulse. The effect of this offset on various quantities can be determined both in simulations and experiment. Comparing the two gives the optimal DC offset needed to compensate the error. In simulation, we compute the time evolution operator which lets us know how big leakage and phase errors are. In an experiment, leakage can be measured by qubit population and phases are accessible with Ramesy measurements.

\begin{figure} \centering
 \includegraphics[width=0.48\textwidth]{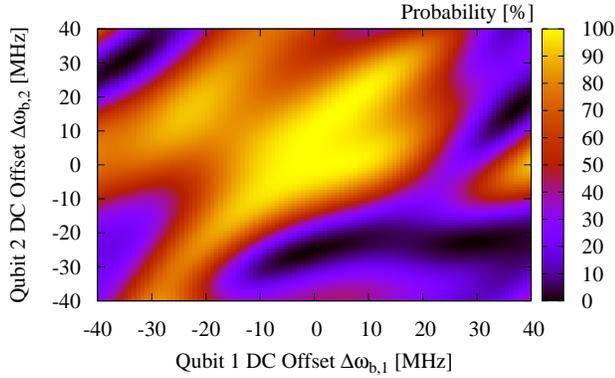}
 \caption{Scan of $|U_{10,10}|^2$ as function of the calibration error. Away from the resonance point $(\Delta\omega_{\text{b},1},\Delta\omega_{\text{b},2})=(0,0)$ leakage starts to manifest itself as a decrease in $|U_{10,10}|^2$. The many features in the plot allow for a good correction of the error.  \label{Fig:ErrBusPop}}
\end{figure}

We illustrate this first with qubit population by scanning $(\Delta\omega_{\text{b},1},\Delta\omega_{\text{b},2})$ and computing the population of qubit one after the gate. Figure \ref{Fig:ErrBusPop} shows the magnitude of $[\hat U]_{10,10}$. It is the entry of the time evolution operator quantifying population transfer from state $\ket{10,0}$ to itself. For an ideal CZ $|[\hat U]_{10,10}|^2=1$, however when changing the DC offset this value decreases. The strong effect of the error is thus used to our advantage since the many features in the $(\Delta\omega_{\text{b},1},\Delta\omega_{\text{b},2})$ error landscape allow an easy comparison between experiment and simulation. Similar data could be obtained with an experiment, comparing it to the simulation would give the DC offset needed to correct the errors. 

\subsubsection{Timing Errors}
\begin{figure}[htbp!] \centering \hfill
 \subfigure[\label{Fig:TimingErrSketch}]{\includegraphics[width=0.48\textwidth]{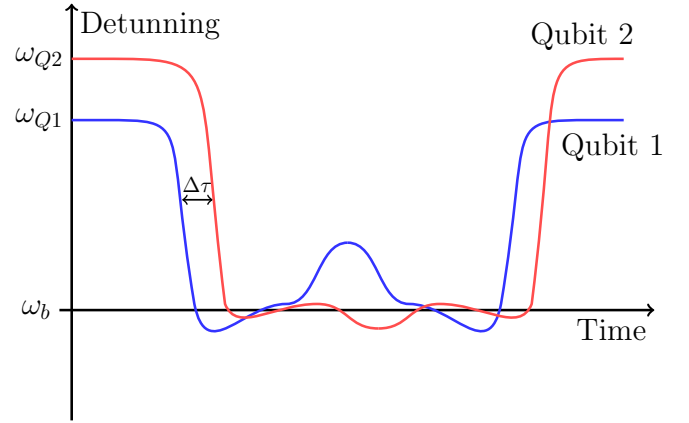}} \hfill
 \subfigure[\label{Fig:TimingErrFid}]{\includegraphics[width=0.48\textwidth]{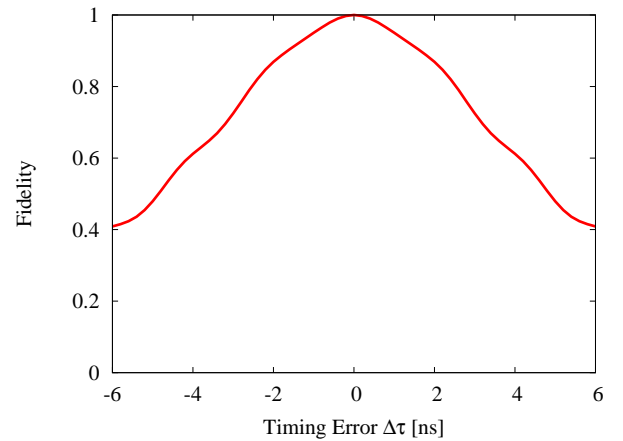}} \hfill
 \caption{Relative timing errors. \subref{Fig:TimingErrSketch} Sketch of pulse relative timing errors. Both pulses have the intended shape but are offset in time by an amount $\Delta\tau$. \subref{Fig:TimingErrFid} Degradation of the fidelity as function of the timing error.}
 \end{figure}

Another error source is the relative timing between the two pulses. This arises if the wires taking pulse one from the AWG to qubit one differ in length from those to qubit two. Pulses offset in time by $\Delta\tau$, as sketched in Fig. \ref{Fig:TimingErrSketch}, lose their fidelity as shown by Fig. \ref{Fig:TimingErrFid}. This error can be removed by introducing a time shift between the pulses and scanning the time shift until leakage/fidelity is minimized/optimized. As seen from Fig. \ref{Fig:TimingErrFid} the relative timing between the pulses should be accurate to within $\approx100~\mathrm{ps}$.

\subsubsection{Hamiltonian Parameter Errors}
\begin{figure}[htbp!] \centering
 \includegraphics[width=0.48\textwidth]{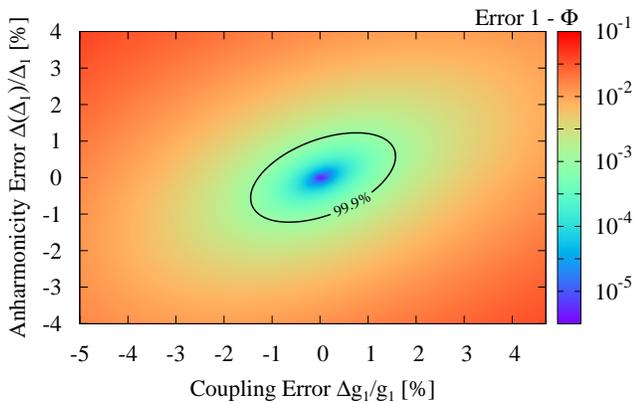}
 \caption{Degradation in fidelity due to errors in the parameters of the Hamiltonian. The pulse was optimized with a target error of $10^{-5}$. If a fidelity of $99.9\%$ is sufficient, the intrinsic robustness of the pulse can support errors of up to $1.5\%$ in coupling strength and $1.2\%$ in anharmonicity. \label{Fig:Couple_anharm_Err}}
\end{figure}

Gradient ascent engineered pulses rely on knowing the Hamiltonian to optimize the pulse. However the parameters entering the model need to be measured and thus come with some amount of uncertainty and error. The pulses designed with these parameters will perform sub-optimally. In the Qubit-Bus-Qubit system, there are four parameters that are susceptible to these error: $\{\Delta_1,\Delta_2,g_1,g_2\}$. For instance, Fig. \ref{Fig:Couple_anharm_Err} shows fidelity degradation as function of errors on the coupling strength and anharmonicity of qubit 1. The pulse was optimized to have a target error of $10^{-5}$. If a pulse fidelity of $99.9\%$ is sufficient, the intrinsic pulse robustness, i.e. $\partial \Phi/\partial g_k\approx0$ and $\partial \Phi/\partial \Delta_k\approx0$ allows us to tolerate and error of up to $1.5\%$ in coupling strength and $1.2\%$ in anharmonicity.

\section{Conclusion}
We develop fast pulses implementing an entangling gate, the CZ,
between two qutrits through a bus. These demonstrate a factor 2 speed up in CZ gates as
well as the possibility to reach arbitrary intrinsic gate fidelity as
long as the gate time is above the quantum speed limit. It turns out
that the optimal pulses break the symmetry of the target gate make
active use of non-computational excited states. We have also
shown how errors arising form realistic experimental conditions can be
negated.

This research was funded by the Office of the Director of National Intelligence (ODNI), Intelligence Advanced Research Projects Activity (IARPA), through the Army Research Office.  All statements of fact, opinion or conclusions contained herein are those of the authors and should not be construed as representing the official views or policies of IARPA, the ODNI, or the U.S. Government.

We acknowledge continued collaboration with D.~Sank and J.M.~Martinis
as well as useful discussions with E.J.~Pritchett, S.T.~Merkel, and
M.R.~Geller. We thank Christine Ridder  for the implementation of BFGS. This work was supported by IARPA through the MQCO and the
EU through the SCALEQIT program.

\bibliography{QC_PublicationDataBase_2}{}

\end{document}